\documentclass[conference]{IEEEtran}
\IEEEoverridecommandlockouts
\pdfoutput=1
\usepackage{cite}
\usepackage{amsmath,amssymb,amsfonts}
\usepackage{algorithmic}
\usepackage{graphicx}
\usepackage{epstopdf}
\usepackage{textcomp}
\usepackage{xcolor}
\usepackage{balance}
\usepackage{placeins}
\usepackage[hidelinks]{hyperref}
\def\BibTeX{{\rm B\kern-.05em{\sc i\kern-.025em b}\kern-.08em
T\kern-.1667em\lower.7ex\hbox{E}\kern-.125emX}}

\usepackage{fancyhdr}

\fancypagestyle{firstpage}{% Page style for first page
  \fancyhf{}% Clear header/footer
  \fancyhead[C]{This is the authors' version of the paper. The final paper appears in the proceedings of the 15th International Symposium on Wireless Communication Systems (ISWCS '19), Oulu, Finland, 27-30 August 2019.}% Header
  \fancyfoot[C]{~\thepage~}% Footer
}
\begin{document}

\title{A Performance Evaluation Tool for Drone Communications in 4G Cellular Networks
\thanks{This work is funded by the security research programme KIRAS of the Federal Ministry for Transport, Innovation, and Technology (bmvit), Austria under grant agreement n.~854747 (WatchDog). The work of C.~Bettstetter is part of the Karl Popper Kolleg on networked autonomous aerial vehicles.}
}
\author{
\IEEEauthorblockN{Christian Raffelsberger$^{*}$, Raheeb Muzaffar$^{*}$, and Christian Bettstetter$^{*+}$}
\IEEEauthorblockA{$^{*}$Lakeside Labs GmbH, Klagenfurt, Austria\\
$^{+}$University of Klagenfurt, Institute of Networked and Embedded Systems, Klagenfurt, Austria\\ \{lastname\}@lakeside-labs.com}\vspace{-0.9cm}
}
\maketitle
\thispagestyle{firstpage}% firstpage page style for first page

\begin{abstract}
We introduce a measurement tool for the performance evaluation of wireless communications with drones over cellular networks. The Android software records various LTE parameters, evaluates the TCP and UDP throughput, and tracks the GPS position. Example measurement results are~presented. 
\end{abstract}


\section{Introduction}
The commercial and industrial market for small-drone appli-\\cations is rapidly growing and provides opportunities in terms of surveillance, monitoring, transport, emergency management, and agriculture, among many other domains. In terms of wireless connectivity, off-the-shelf drones are usually equipped with Wi-Fi. Whereas this is sufficient for applications with low coverage requirements, cellular networks could provide wide area coverage, which is especially important for autonomous flights extending beyond line of sight~(LoS). However, today's cellular networks\,---\,including Long Term Evolution Advanced~(\mbox{LTE-A})\,---\,are not optimized for aerial devices~\cite{BerghCP16}. For example, the base station antennas are typically tilted downwards to serve users on the ground rather than flying devices. Technical enhancements are necessary to optimize the connectivity of drones, e.g., to manage interference and handovers~\cite{lin2018sky}. Upcoming cellular systems\,---\,with advanced beamforming, beam steering, and full dimensional multi-input multi-output~(FD-MIMO)\,---\,may alleviate shortcomings to provide better aerial connectivity~\cite{tadayon2016inflight,muruganathan2018overview}.

Several experimental studies have explored drone communication with Wi-Fi (see \cite{gu2015airborne,6815903,yanmaz2011channel,khuwaja2018survey}) but few relate to cellular-based drone communication (see  \cite{lin2018sky,BerghCP16,yang2018telecom,sundqvist,8641420}). In order to study cellular-connected drones, the research community needs an evaluation tool to analyze the performance in current and forthcoming 3rd Generation Partnership Project (3GPP) releases and to design drone applications accordingly. This paper introduces such a tool, namely the {\it cellular drone measurement tool}~(CDMT). It integrates several important features to evaluate cellular communication in aerial networks, such as monitoring of signal strength and cell information, assessing Transmission Control Protocol (TCP) and User Datagram Protocol (UDP) performance, and tracking Global Positioning System (GPS) coordinates. Some results obtained with CDMT are presented in this paper. We demonstrate the tool in action by flying an AscTec Pelican drone with an Android smartphone. In our experiment, we observe some ping-pong handovers and recognize that the downlink~(DL) throughput is higher on the ground, whereas the uplink~(UL) throughput is higher in the air. More results on handovers and throughput using CDMT are given in \cite{fakhreddine19:dronet} and~\cite{hayat19}. 

The remainder of the paper is organized as follows: Section~II provides an overview of available LTE measurement tools, before Section~III describes LTE signal metrics. Section~IV introduces CDMT and presents its features. Section~V shows some example experiments, and finally Section~VI concludes the paper.

\begin{figure*}[!t]
\begin{center}
\scriptsize
\begin{tabular}{ccc}
\centering 
\vspace{0.2cm}
$\vcenter{\hbox{\includegraphics[width=0.165\textwidth]{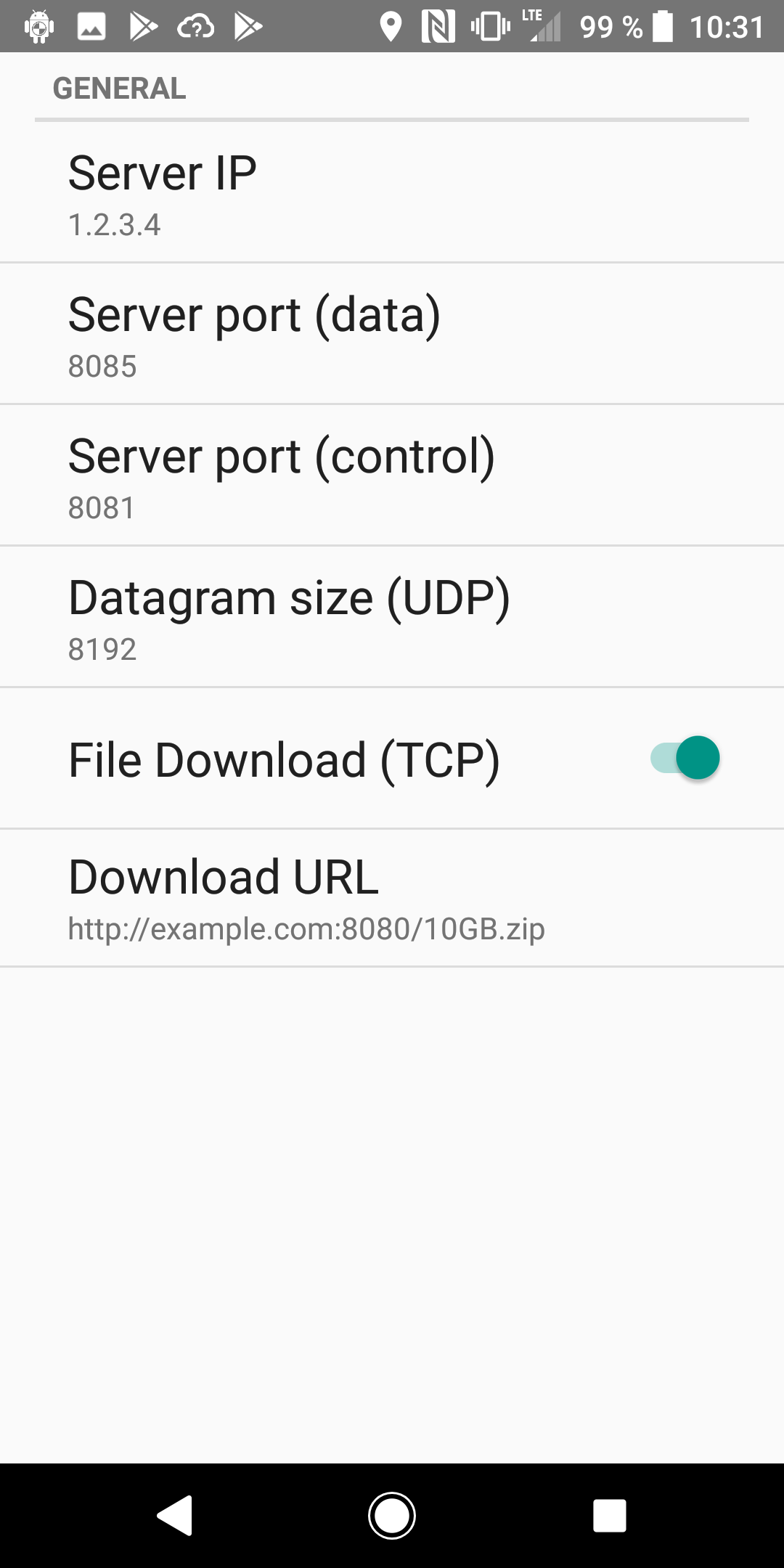}}}$ &
$\vcenter{\hbox{\includegraphics[width=0.165\textwidth]{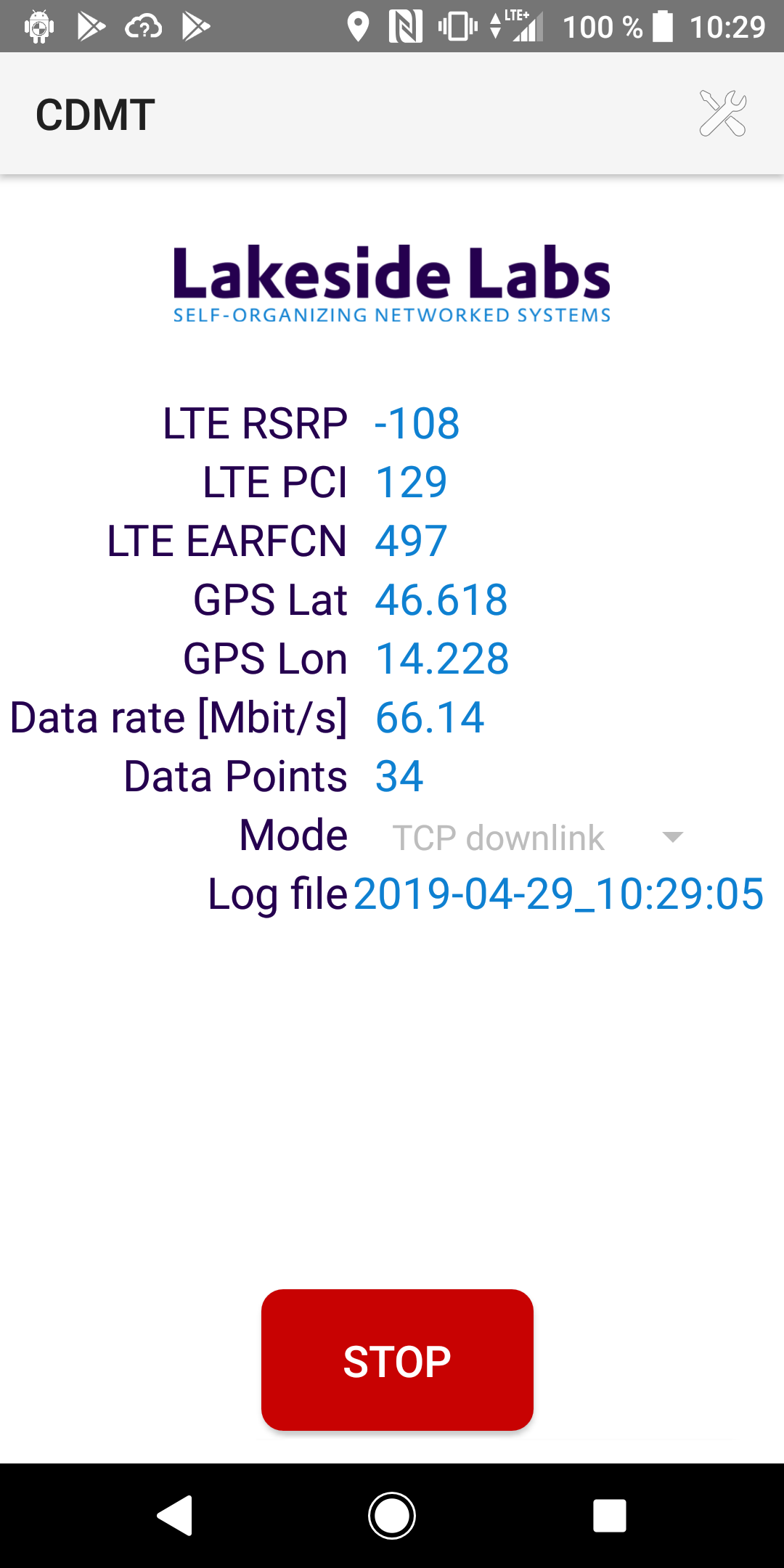}}}$ &
$\vcenter{\hbox{\includegraphics[width=0.165\textwidth]{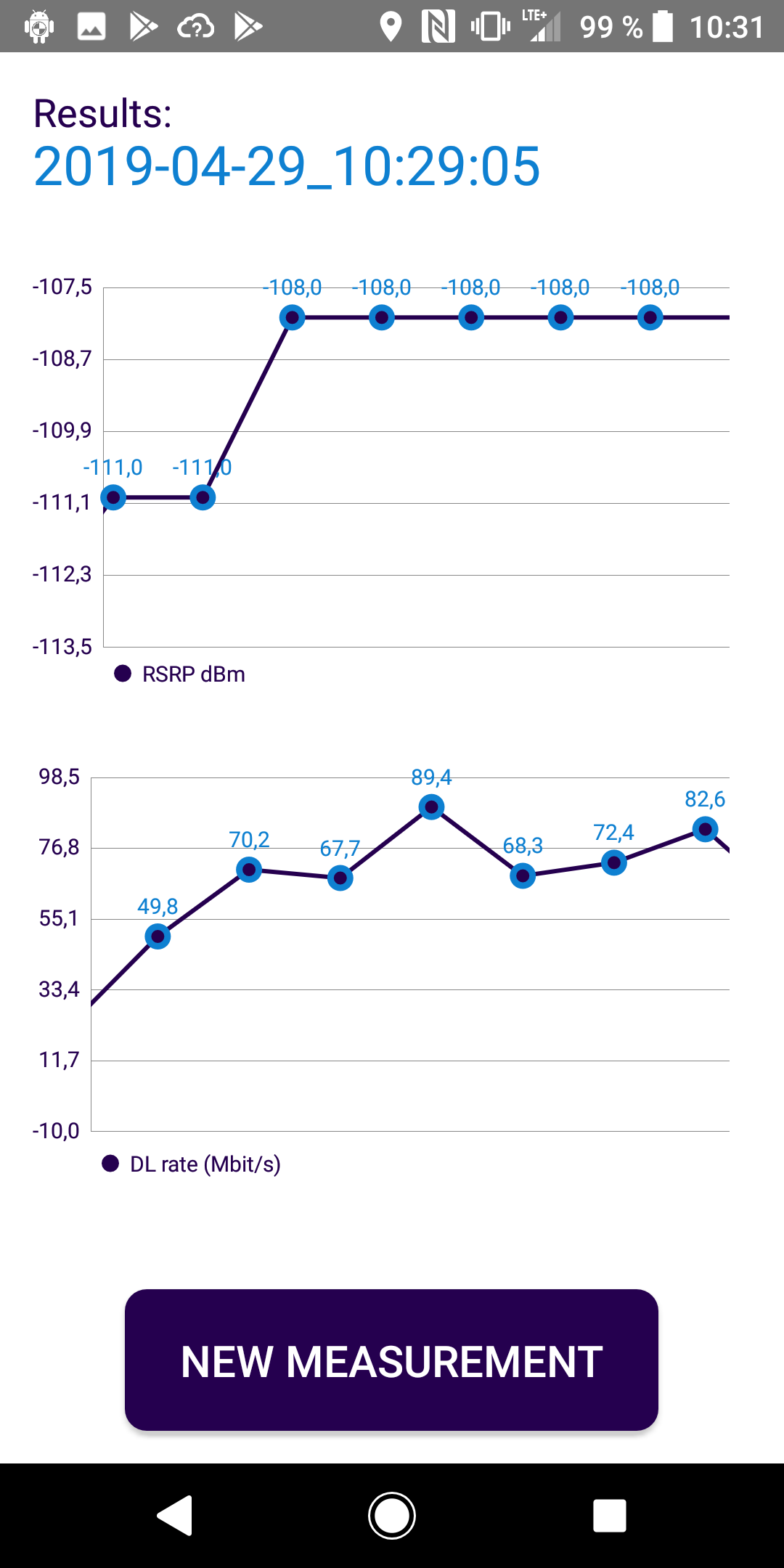}}}$ \\
\vspace{-0.2cm}
%\hspace{-0.95cm}(a) & \hspace{1.4cm}(b)& \hspace{1.4cm}(c) \\
(a)~Configuration & (b)~Running experiment & (c)~Results overview\\
\end{tabular}
\renewcommand{\arraystretch}{1}
\end{center}
\caption{CDMT screenshots} 
\label{fig:cdmt-screenshots}
\end{figure*}

\section{Related Tools}
While some LTE measurement tools with various features are freely available, to the best of our knowledge, none of them integrates all of the parameters supported by CDMT. Moreover, they usually focus exclusively on TCP as the transport layer protocol. Commercial solutions offering similar parameters such as CDMT are expensive and often reside within the operator's network~\cite{xcal,shafiq2014understanding}.
The performance and power characteristics of LTE, 3G, and Wi-Fi are analyzed in~\cite{huang2012close}. The  Android tool 4GTest~\cite{MobiPerf} is employed to characterize the LTE network performance. It uses a client-server model with multiple servers deployed at different locations and allows users to switch between network types to capture performance measurements. 
A model to characterize the round trip time and throughput over the signal strength is derived from measurements with a customized measurement tool in \cite{albaladejo2016measurement}. The tool uses a client-server model that records the UDP performance in the DL. It is not publicly available and can track only a few parameters of cellular networks. 

The Nemo Handy Handheld Measurement Solution \cite{nemo} is a commercial Android-based network testing application. It provides many relevant parameters, including connection setup delay, download time, time-to-connect delay, throughput, reference signal received power~(RSRP), reference signal received quality~(RSRQ), latitude, longitude, altitude, and information about neighboring cells. 
This tool has been used for LTE measurements with a drone in \cite{sundqvist}. It was observed that the signal strength in that scenario increases from $-93$~dBm to $-74$~dBm until the drone reaches an altitude of $34$~m, before subsequently declining again with rising flight height. 
Another measurement campaign with a drone and Nemo was carried out to analyze LTE signals (RSRP and signal-to-interference ratio) in dense environments up to an altitude of $350$~m~\cite{8641420}. Here, the signal strength increases as the drone clears the height of the buildings due to LoS links with the base stations. However, higher path loss and interference from neighboring cells adversely affects the received signal power.

The study \cite{huang2013depth} addresses LTE and Wi-Fi and analyzes whether transport protocols should select the best network or better use multipath TCP. The ``Cell vs WiFi"~\cite{cellvsWiFi} app measures and compares the performance of LTE and Wi-Fi on Android devices. The app is currently unavailable.

In conclusion, available tools are either proprietary and not freely available or designed in simplicity, not capturing all important parameters necessary to analyze cellular-based drone communication. Thus, we developed a measurement tool that supports both TCP and UDP and captures information about signal strength, frequency bands used, and neighboring~cells. 

\section{LTE Signal Metrics}
The three metrics related to LTE signal strength available on Android devices are RSRP, RSRQ, and RSSNR (reference signal signal-to-noise ratio)~\cite{cainey2014modelling}. First, RSRP provides the average power received by the resource element carrying the reference signal in any symbol. It ranges from $-44$~dBm to $-140$~dBm\cite{etsi:ts127007}. It is primarily used for cell selection and handover. However, a high throughput can be observed even with low RSRP.
Second, RSRQ provides additional information about when to perform a handover. It indicates the quality of the reference signal with typical values ranging between $-19.5$~dB (worst) and $-3$~dB (best)\cite{etsi:ts127007}. 
Third, RSSNR measures the signal-to-noise ratio of the received signal. It is used to assess the the signal quality and is an additional parameter to make handover decisions. Typical values range from $30$~dB (best) to $-20$~dB~(worst)\cite{etsi:ts136101}. 

The UE sends these parameters to the serving base station (eNodeB) as a measurement report. Acoordingly, the eNodeB makes the handover decision based on the reports received~\cite{racz2007handover}.

\section{Cellular Drone Measurement Tool}

CDMT is a performance evaluation tool (see Fig.~\ref{fig:cdmt-screenshots}) developed for the Android platform to be used for aerial devices connected with 4G cellular networks. It is available~at
\begin{center}
{\tt https://www.lakeside-labs.com/cdmt}
\end{center}
\noindent and can be utilized for academic research. The CMDT records LTE parameters such as RSRP, RSRQ, RSSNR, cell signal quality~(CSQ), serving physical cell identity~(PCI), channel quality indicator~(CQI), E-UTRA absolute radio frequency channel number~(EARFCN), neighboring cell information including PCI, EARFCN, RSRP, and RSRQ. It supports throughput measurements for TCP and UDP for DL and UL, packet delay for UDP, and it logs GPS information (time, latitude, longitude, altitude, number of available satellites, speed, and acceleration). All parameters are logged at a rate of~$1$~Hz.

In order to measure throughput and round trip time (RTT), CDMT exploits a client-server model, where the client application is executed on Android. On the client side, the server network configuration (e.g., IP address and ports) and\,---\,depending on the type of measurement\,---\,some further parameters (e.g., UDP segment size) have to be configured (Fig.~\ref{fig:cdmt-screenshots}(a)). Once the parameters are configured and the server application is running, measurements can be recorded (Fig.~\ref{fig:cdmt-screenshots}(b)). Graphical representations of RSRP and throughput measurements are shown once the recording has been finished (Fig.~\ref{fig:cdmt-screenshots}(c)).

The server application is written in Java. It comprises two modules: the control module starts and stops measurements and reports results to the client, while the data module sends or receives data via TCP and UDP sockets. The UDP data rate is measured on the client for downlink tests and on the server for uplink tests. In the latter case, the server reports the data rate achieved to the client, which stores it in a log file. The TCP data rate is always calculated on the client. 

It is assumed that the server is accessible via a public Internet Protocol (IP) address. Hence, we can always perform tests for the TCP uplink, TCP downlink, and UDP uplink. UDP downlink tests may not be available depending on the Network Address Translation (NAT) configuration of the mobile operator. CDMT only provides a simple NAT traversal technique: the client first sends a UDP packet to the server to create a mapping at the network address translator, and the server then uses the external IP address and port combination of the received packet as an endpoint for the UDP downlink. However, this mechanism does not work if the NAT changes the port mapping during the measurement. Since this was the case for our mobile operator, we selected an Access Point Name (APN) setting that does not use carrier NAT but rather a public IP address configuration at the end device instead.

The TCP throughput is measured by downloading or uploading a random stream of data created at the server or client, respectively. Alternatively, the TCP downlink test can be made using the Hypertext Transfer Protocol (HTTP)  via a Uniform Resource Locator (URL), pointing to a file that is then (repeatedly) downloaded by the client until the measurement is manually stopped. In this configuration, the custom measurement server is not involved. The UDP throughput is measured by sending data packets of configurable size. The first four bytes contain the sending timestamp (to calculate the transmission delay) and the remaining bytes are chosen randomly. This requires synchronized  clocks at the server and the Android device. The calculated delay is reported back using the TCP control socket.

The user starts and terminates a test manually. Measurement values are recorded every second and stored on the client platform for offline analysis. Most existing tools do not support such recording but rather only provide averaged values at the~end. 

\section{Experimental Study}
This section describes an experimental study that showcases the possibilities of the CDMT. Further experiments with CDMT\,---\,along with more comprehensive results on throughput and handover rates\,---\,are presented in~\cite{fakhreddine19:dronet} and~\cite{hayat19}. 

\begin{figure}[t]
\centering
\includegraphics[width=2.9in]{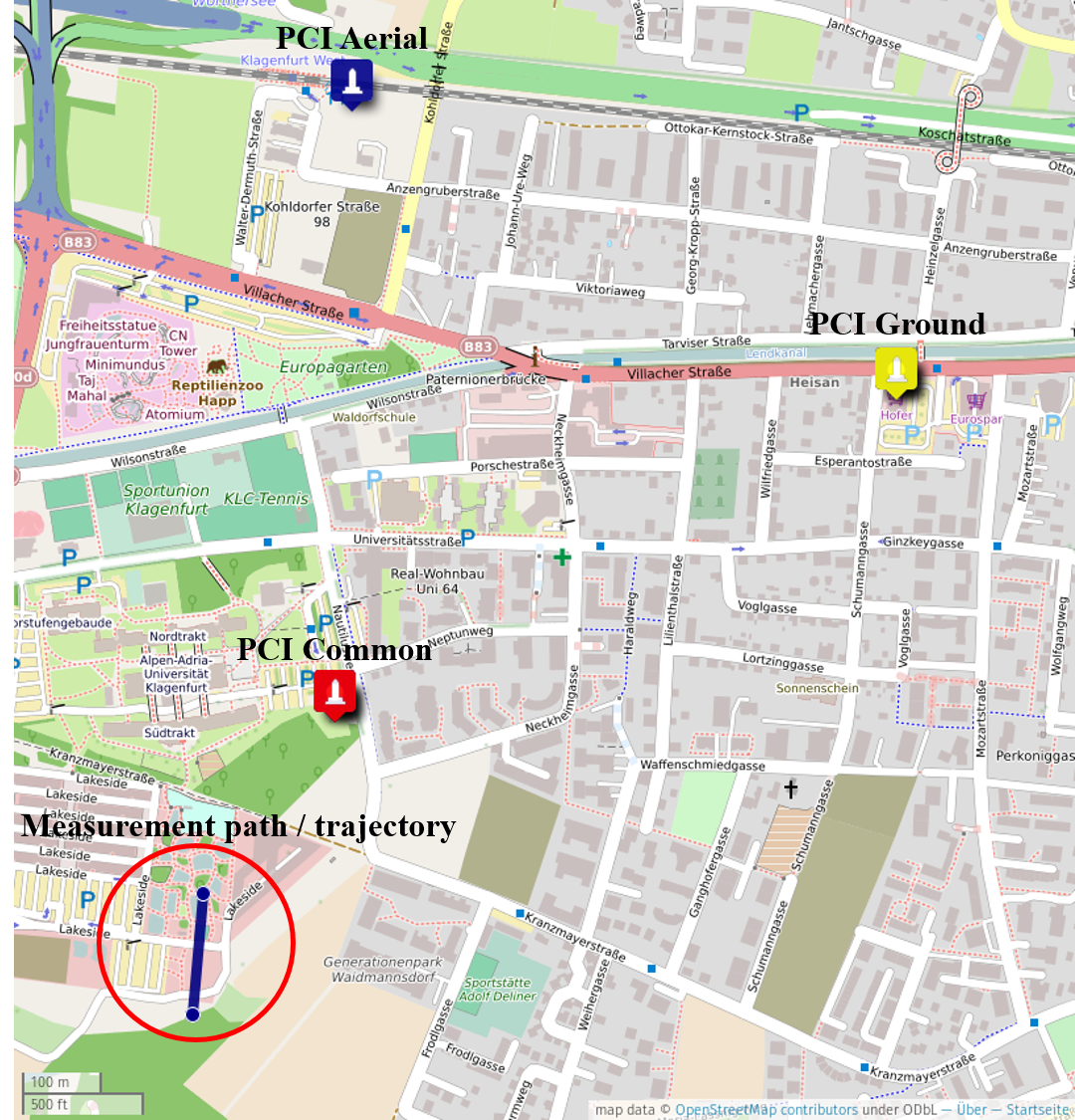}
\caption{Measurement path and trajectory with the location of serving PCIs}\vspace{-0.2cm}
\label{fig:trajectory}
\end{figure}

\subsubsection{Setup}
A Sony Xperia H8216 smartphone running Android~$8.0$ is used. It has two quad-core processors with $4$~GB RAM and the Qualcomm Snapdragon~845 chipset supporting LTE carrier aggregation~(CA). It is mounted on an AscTec Pelican drone, which can carry a  payload of $650$~g with a flight time of about $16$~minutes. We evaluate the throughput over \mbox{LTE-A} (3GPP Release~13) with ground and aerial measurements. CA is activated in the DL on four frequency bands (LTE800, LTE1800, LTE2100, and LTE2600), and no CA is used in the UL. The maximum antenna output power is~$46$~dBm.

For UDP experiments, a packet size of $8,\!192$~bytes is chosen. TCP downlink tests use the file download mode (i.e., a $1$~GB file is repeatedly downloaded from a public server). All other experiments (TCP UL, UDP UL and DL) are performed by configuring CDMT to connect to a measurement server located at our lab and running the aforementioned Java module. Measurements are performed as line tests from a reference point to a distance of $150$~m. The drone is set to fly at an altitude of $50$~m. Fig.~\ref{fig:trajectory} shows the flight path trajectory and the locations of the serving PCIs.

\subsubsection{Throughput}

\begin{figure}[t]
\centering
\vspace{-0.4cm}
\includegraphics[trim=1cm 0cm 1.5cm 0cm, width=2.9in]{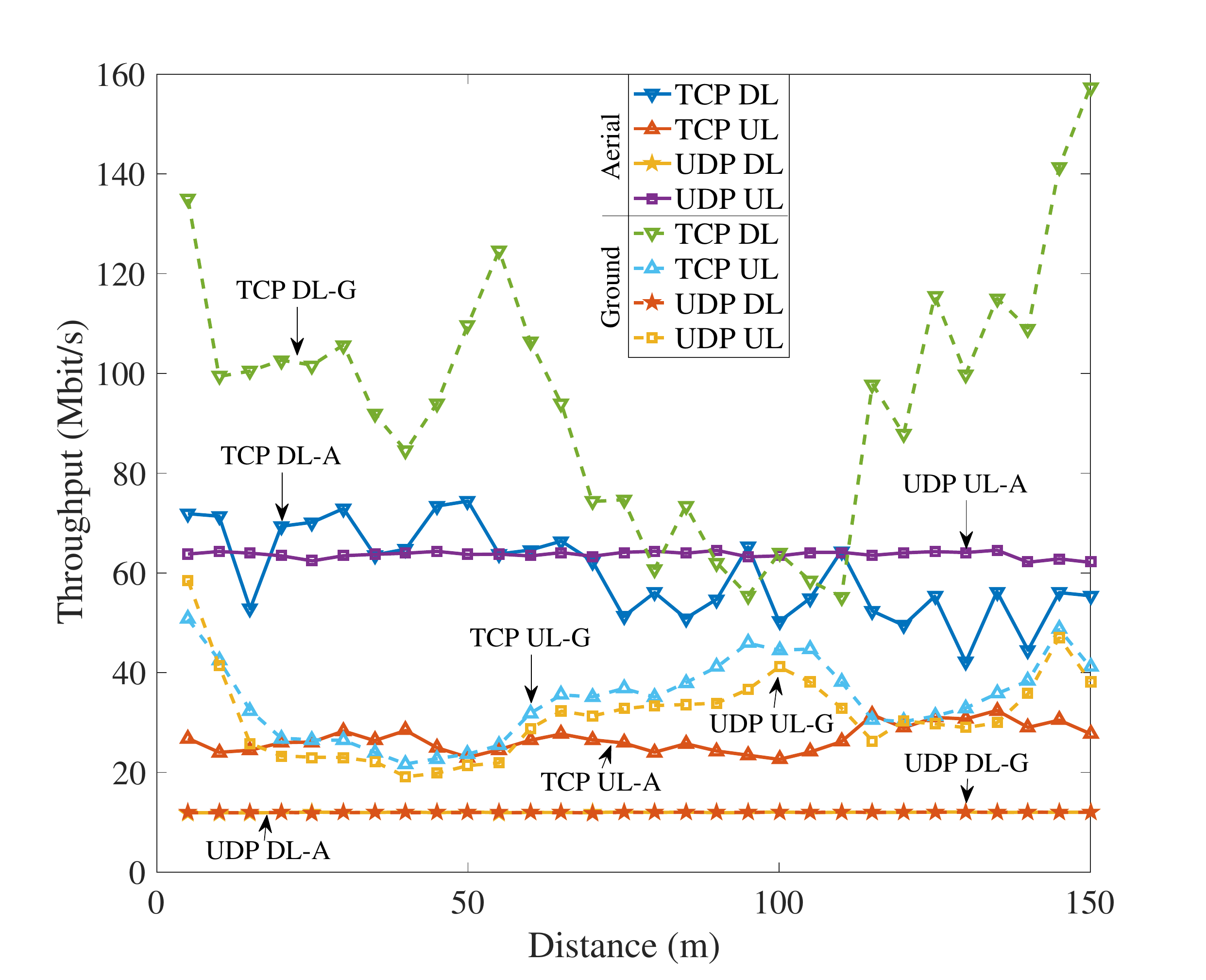}\vspace{-0.2cm}
\caption{Average throughput on the ground and $50$~m in the air}\label{fig:throughput}
\end{figure}

Fig.~\ref{fig:throughput} shows the measured throughput over flight distance. Eight curves are given for the throughput in the DL (eNodeB to UE) and UL (vice versa) using either TCP or UDP for both aerial and ground measurements. The presented curves are the mean of three experimental runs. It can be observed that the TCP throughput is higher on the ground than in the air (for both DL and UL), while the reverse is true for the UDP throughput in the UL. The UDP throughput in the DL on the ground and in the air are about the same. 

\begin{figure}[t!]
\centering
\vspace{-0.4cm}
\includegraphics[trim=1cm 0cm 1.5cm 0cm, width=2.9in]{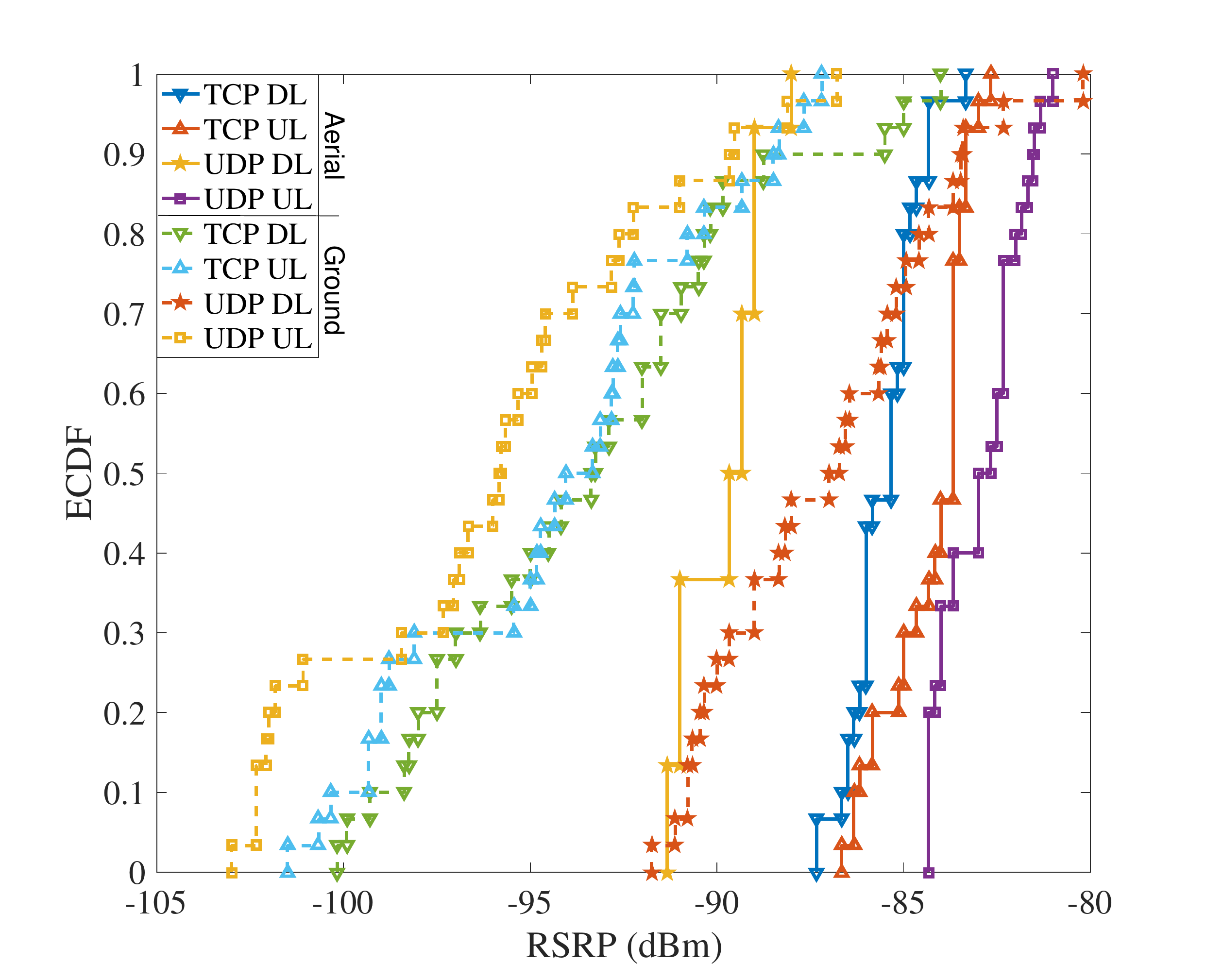}\vspace{-0.2cm}
\caption{Empirical cumulative distribution of RSRP on the ground and $50$~m in the air}
\label{fig:RSRP_ALL}
\end{figure}

The behavior of the TCP throughput can be explained by the fact that the RSRP is always better on the ground (Fig.~\ref{fig:RSRP_ALL}). For UDP, the RSRP on the ground is much worse than in the air. However, an actual relationship between RSRP and throughput cannot be established. 

\subsubsection{Handovers}

In the aerial scenario, in the TCP DL we observed a ping-pong handover from PCI 130 to 388 and back to 130 once in one run. No handover was observed during the other two runs as the UE was always connected to PCI 130 or 92, respectively. 
In the ground scenario, we observed several ping-pong handovers: in the three TCP DL runs, changes were from PCI 263 to 56 and back to 263, from 130 to 295 and back to 130, and from 92 to 109 and back to 92.
In the TCP UL, handovers happened from PCI 263 to 56 and back to 263 (in two runs) and from 92 to 109 and back to 92 (in one run). No handover was observed in the TCP aerial UL. The PCIs selected were 92, 263, and 130 for the three runs.

In the UDP DL, no handover was observed on the ground nor in the air. The UE connected to PCI 263 in the air and to PCI 359 on the ground. 
In the UDP UL, the UE remained connected to PCI 263 in all three aerial experiments and performed a ping-pong handover from PCI 263 to 56 and back to 263 in all three ground experiments. 

\section{Conclusions and Outlook}
The cellular drones measurement tool~(CDMT) targets aerial devices connected to 4G  networks. It can be used to record LTE performance parameters available on Android platforms, including TCP and UDP throughput, and supports tracking via GPS. We demonstrated its feasibility with example results from a measurement campaign over \mbox{LTE-A}. These results show better TCP downlink and uplink throughput on the ground and higher UDP uplink throughput in the air. Accompanying and future work includes comprehensive experimental evaluations at different flight heights and an adaptation of CDMT for 5G~networks.

\bibliographystyle{IEEEtran}
\balance
\bibliography{ISWCS_2019}

\end{document}